\documentclass{article}

\PassOptionsToPackage{numbers,compress}{natbib}

\usepackage{comment}
\usepackage{enumitem}
\usepackage{multirow}
\usepackage{siunitx}
\usepackage{graphicx}
\setlist{nolistsep}

\usepackage[final]{neurips_2023_ml4ps}

\usepackage[utf8]{inputenc} 
\usepackage[T1]{fontenc}    
\usepackage{hyperref}       
\usepackage{url}            
\usepackage{booktabs}       
\usepackage{amsfonts}       
\usepackage{nicefrac}       
\usepackage{microtype}      
\usepackage{xcolor}         

\usepackage{wrapfig}

\title{AI ensemble for signal detection of 
higher order gravitational wave modes of 
quasi-circular, spinning, non-precessing binary 
black hole mergers}

\author{%
  Minyang Tian \\
  Data Science and Learning Division, \\ 
  Argonne National Laboratory, \\
  Lemont, Illinois 60439, USA\\
  Department of Physics \& NCSA, \\ 
  University of Illinois at Urbana-Champaign, \\
  Urbana, Illinois 61801, USA\\
  \texttt{mtian8@illinois.edu} \\
   \And
   E. A. Huerta\\
  Data Science and Learning Division, \\ 
  Argonne National Laboratory, \\
  Lemont, Illinois 60439, USA\\
  Department of Computer Science \\
  The University of Chicago, \\ 
  Chicago, Illinois 60637, USA\\
  \texttt{elihu@\{anl.gov,uchicago.edu\}} \\
  \And
   Huihuo Zheng \\
  Leadership Computing Facility, \\ 
  Argonne National Laboratory, \\
  Lemont, Illinois 60439, USA\\
  \texttt{huihuo.zheng@anl.gov} \\
}

\begin{document}

\maketitle

\begin{abstract}
We introduce spatiotemporal-graph 
models that concurrently process data  
from the twin advanced LIGO detectors and the advanced 
Virgo detector. We trained these AI classifiers 
with 2.4 million \texttt{IMRPhenomXPHM} 
waveforms that describe quasi-circular, spinning, 
non-precessing binary black 
hole mergers with component masses \(m_{\{1,2\}}\in[3M_\odot, 50 
M_\odot]\), and individual spins 
\(s^z_{\{1,2\}}\in[-0.9, 0.9]\); and which include the 
\((\ell, |m|) = \{(2, 2), (2, 1), (3, 3), (3, 2), (4, 4)\)\} 
modes, and mode mixing effects in the 
\(\ell = 3, |m| = 2\) harmonics. We trained these 
AI classifiers within 22 hours using distributed 
training over 96 NVIDIA V100 GPUs in the Summit 
supercomputer. We then used transfer learning to 
create AI predictors that estimate the total mass 
of potential binary black holes identified by 
all AI classifiers in the ensemble. 
We used this ensemble, 3 classifiers for signal detection 
and 2 total mass predictors, to process a year-long 
test set in which we injected 300,000 signals. 
This year-long test set was processed within 
5.19 minutes using 1024 NVIDIA A100 GPUs in the Polaris 
supercomputer (for AI inference) 
and 128 CPU nodes in the ThetaKNL supercomputer 
(for post-processing of noise triggers), housed 
at the Argonne Leadership Computing Facility. 
These studies indicate that our AI ensemble provides 
state-of-the-art signal detection accuracy, and reports 
2 misclassifications for every year of searched data. 
This is the first AI ensemble designed to search for and 
find higher order gravitational wave mode signals.
\end{abstract}

\section{Introduction} 

Artificial Intelligence (AI) applications have led to 
remarkable breakthroughs in science, engineering,  
industry, and tech during the last 
decade~\cite{DeepLearning}. Novel AI 
applications are now being explored in earnest to 
address contemporary scientific grand challenges, 
as well as to provide a platform to elevate human 
insight across disciplines~\cite{2022NatRP761K,Davies2021AdvancingMB}. Gravitational wave 
astrophysics is part of this 
revolution. For several years now, an ever growing, 
international community of researchers has 
been developing novel AI tools to maximize the science 
reach of gravitational wave astrophysics~\cite{Nat_Rev_2019_Huerta,cuoco_review,huerta_book,huerta_nat_ast}. In this article, we contribute to the development 
of AI to search for and find higher order 
gravitational wave modes emitted by quasi-circular, 
spinning, non-precessing, stellar mass binary black hole 
mergers. 

\paragraph{Assumptions.} We consider a three detector network 
    comprising the advanced LIGO
    Livingston (L) and Hanford (H) detectors, and advanced 
    Virgo (V).
We train, validate and test our AI models 
    using modeled waveforms that include 
    higher order gravitational wave modes, and 
    mode mixing effects. We use colored Gaussian noise throughout 
    this analysis to benchmark this approach. In 
    future work we will extend this analysis using 
    real gravitational wave noise. 
To model the sensitivity of our proposed three detector network, we use the following target power spectral density (PSD) noise curves: aligo\_O4high.tx for the advanced LIGO detectors, and avirgo\_O5low\_NEW.txt for 
advanced Virgo~\cite{Abbott2020,pdsnoise}. 
    We do this to consider gravitational wave detectors with 
    comparable sensitivity.

\paragraph{Claims.}
To the best of our knowledge, we present the first AI models in the literature 
    trained for signal detection of higher order 
    gravitational wave modes emitted by quasi-circular, 
    spinning, non-precessing binary black hole mergers.
We quantified the performance of these AI models  
    for signal detection by processing a year of coloured Gaussian noise in which we injected 300,000 
    higher order gravitational wave mode signals. 
    Using the Receiver Operating Characteristic 
    Area Under the Curve (ROC AUC), and the Precision 
    Recall Area Under the Curve (PR AUC), we found that 
    our AI classifiers provide state-of-the-art signal 
    detection accuracy, and report 3 false positives over 
    a year's long test set. Once we post-processed 
    noise triggers with 2 AI total mass predictors, we 
    reduced the number of misclassifications to only 2 
    over an entire year of searched data.

\paragraph{Limitations.} Our AI models have been designed 
assuming a three detector network, LHV. We will explore 
larger detector networks in future work. Our AI models 
have been trained, validated and tested using coloured 
Gaussian noise, and target PSDs for the sensitivity of 
the LHV network. This is done to understand the 
performance of our AI models with controlled 
experiments, and to develop the required scientific 
software to handle these large datasets, and to train 
models at scale. In future work we will leverage this 
knowledge and scientific software to develop new AI 
models using gravitational wave data from the 
Gravitational Wave Open Science Center~\cite{KAGRA:2023pio}.

\noindent \textbf{Reproducibility.} We release our 
AI models, along with a tutorial that provides a 
step-by-step guide to use them, in the 
following \texttt{GitHub} repository: 
\url{https://github.com/mtian8/Higher_Order_GW_Spatiotemporal_GNN}.

\section{Methods}

\paragraph{Datasets.} We produced a set of 
3 million waveforms with the 
\texttt{IMRPhenomXPHM} 
approximant~\cite{2021PhRvD.103j4056P}. 
These modeled waveforms are one second long, 
sampled at 4096 Hz, and describe 
quasi-circular, spinning, non-precessing 
binary black hole mergers with component masses 
\(m_{\{1,2\}}\in[3M_\odot, 50 
M_\odot]\), and individual spins 
\(s^z_{\{1,2\}}\in[-0.9, 0.9]\); and which include the 
\((\ell, |m|) = \{(2, 2), (2, 1), (3, 3), (3, 2), (4, 4)\)\} 
modes, and mode mixing effects in the 
\(\ell = 3, |m| = 2\) harmonics. We sampled the 
individual masses and spin components 
using a uniform distribution.
The right ascension and declination 
are sampled uniformly on a solid angle of a sphere. 
The polar angle, $\theta$, covers the range 
\(\theta\in[-\pi/2, \pi/2]\), 
while the orbital inclination is 
sampled using a $\sin(\textrm{inclination})$-distribution. 
The coalescence phase and waveform polarization are 
both uniformly sampled covering the range $[0,2\pi)$. 
We created three non-overlapping sets, namely, 
2.4M waveforms for training, 300k waveforms 
for validation, and 300k waveforms for testing.

\paragraph{Sensitivity of detectors.} We used the PSD aligo\_O4high.txt to model the sensitivity of 
advanced LIGO 
detectors, and avirgo\_O5low\_NEW.txt for 
advanced Virgo~\cite{Abbott2020,pdsnoise}.
We used this approach to ensure that all detectors 
in the array have comparable sensitivity for 
signal detection. 

\paragraph{Data curation.} We used \texttt{PyCBC} 
~\cite{pycbc_library} to produce 
\texttt{IMRPhenomXPHM} 
waveforms, and to produce Gaussian noise. 
Both modeled waveforms and Gaussian noise 
were whitened with the PSDs described above, and 
then added linearly to simulate 
events with a broad range of 
signal-to-noise ratios (SNRs). 
Given the sparsity of gravitational wave observations, we prepared our 
training sets so that 
70\% of samples contain only pure noise (negative samples), while 
30\% contain signals contaminated with noise 
(positive samples). Negative samples 
consist of 1s-long 
segments of pure synthetic noise labeled as 0s. 
To generate positive samples, we truncate 
a whitened signal and consider only the 0.5s before merger. We set the label for positive samples 
to be 1 only during 
the 0.5s before merger, while the rest of the signal 
will be set to 0. This is done because our AI models 
respond sharply to the presence of black hole merger 
signals in the vicinity of the merger event.

\begin{wrapfigure}{r}{0.75\textwidth}
   \centering
   \includegraphics[width=0.75\textwidth]{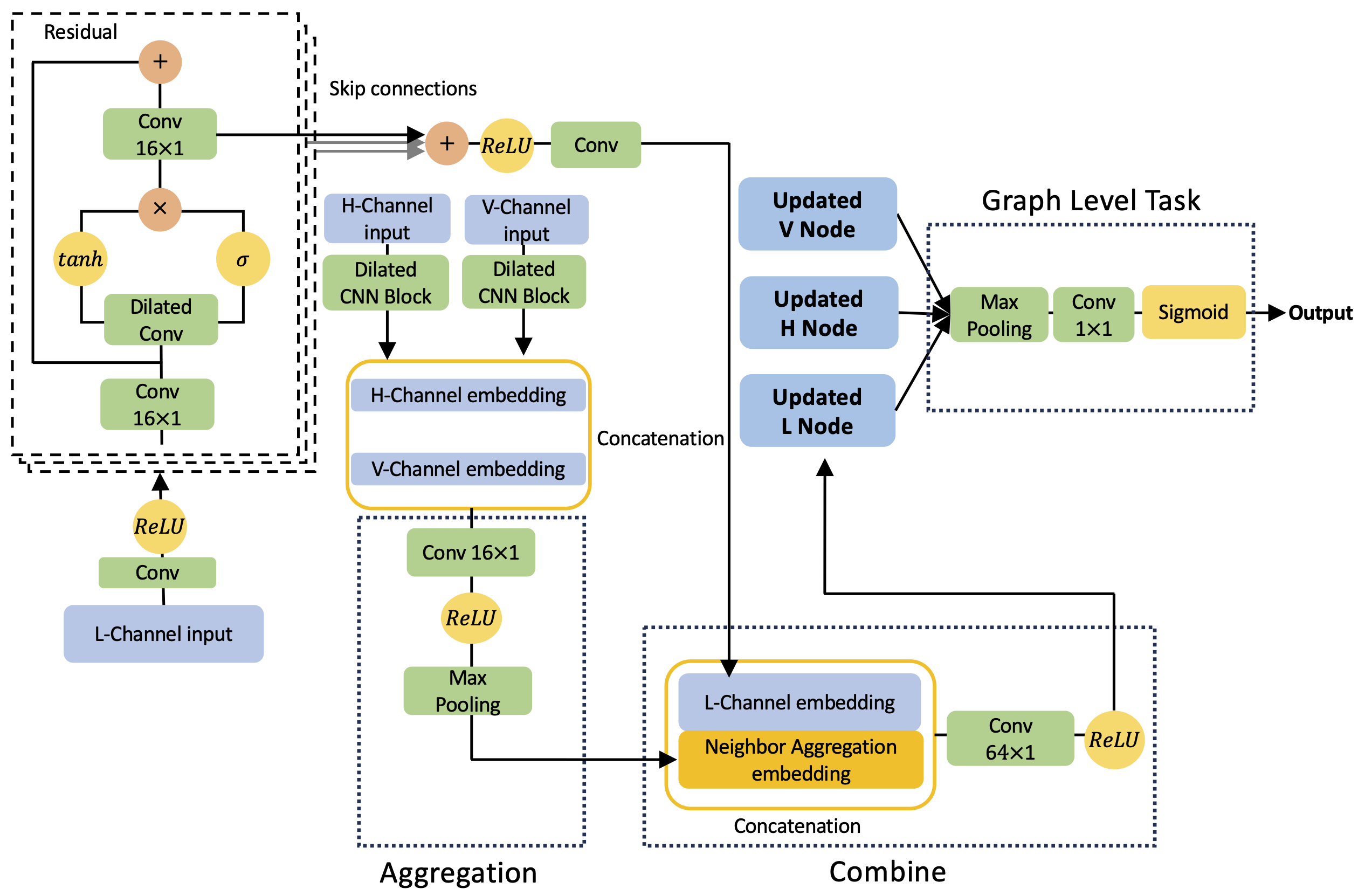}
   \caption{Model structure including the HDCN and GNN blocks.}
   \label{fig:model}
\end{wrapfigure}

\paragraph{AI model architecture.} We consider the 
model architecture introduced 
in Reference~\cite{2023arXiv230615728T}, which 
combines a hybrid dilated 
convolution network (HDCN) ~\cite{2016wavenet} block, and a graph neural 
network (GNN)~\cite{bruna2013spectral, gilmer2017neural, NIPS2017_5dd9db5e, velickovic2017graph, xu2018powerful} block. The HDCN block is used to 
model temporal properties of gravitational wave signals, 
while the GNN block captures geometrical properties of 
the three detector network LHV.  We illustrate the 
model structure in Figure~\ref{fig:model}.

 Each of the three distinct HDCN blocks produces processed embeddings for their respective channels. This approach is critical as it allows the preservation of each detector’s unique information, which is essential for the subsequent GNN block. During this phase, the data traverse through a series of dilated convolution layers, each having varied dilation rates which are used to achieve exponential expansion in the receptive field relative to the number of layers. The output of 3 blocks will be used as GNN node inputs. For every target node, the GNN block performs both an Aggregate and a Combine operation. The layers used for Aggregate and Combine operations are shared among the three nodes. Subsequently, max pooling is employed to amalgamate the embeddings of the three nodes, producing a singular graph-level embedding. Lastly, a Multi-Layer Perceptron (MLP) with sigmoid activation is applied to the graph-level embedding to formulate element-wise predictions. 
 The architecture of the mass estimation model mirrors that of the detection model. The only difference is in the final layer, since we added an additional convolution layer with activation to yield the total mass estimation.

\paragraph{Training methodology.}
We trained our AI models within 22 hrs using 96 
NVIDIA V100 GPUs at the \href{https://www.olcf.ornl.gov/summit/}{Summit} supercomputer 
at Oak Ridge National Laboratory. 
For the purpose of distributed training, we employed \texttt{Horovod}~\cite{sergeev2018horovod}. We used curriculum learning, methodically lowering the SNR to the targeted range. This approach facilitates more efficient and expedited convergence by initially exposing the model to simpler examples. We initiated the learning rate at 1e-3, and halved it if no improvements were observed over three consecutive epochs. To handle our large training sets, 
we used the \texttt{LAMB}~\cite{you2019large} optimizer, 
since it provides adaptivity and efficiency in the training process.

\section{Results}

\noindent \textbf{AI models raw output.} Our AI classifiers produce outputs in the form of element-wise probabilities, indicating the likelihood of each data point being part of a gravitational wave. 

\noindent \textbf{Post-processing analysis.} AI outputs  
require further refinement to produce final detection results,  since any legitimate signal exist within at least half a second, equivalent to 2048 data points. To accomplish this, we employ the \texttt{find\_peaks} algorithm from \texttt{SciPy}~\cite{2020SciPy-NMeth}. This algorithm processes the output from our detection models and pinpoints locations that meet the prerequisites of being at least ${\cal H}$ in height and 2000 in width. Here, ${\cal H}$ represents a modifiable threshold, pivotal 
for computing ROC and PR curves. To achieve a minimal False Positive Rate (FPR), we set a high threshold of 0.999999, ensuring that we sustain a substantial True Positive Rate (TPR) simultaneously. 

\noindent \textbf{AI performance.} To quantify the performance of our AI models, we prepared a year's long test set in which 
we injected 300,000 \texttt{IMRPhenomXPHM} waveforms. 
We first processed this test using the 3 AI 
classifiers within 165 seconds using 1024 A100 NVIDIA 
GPUs in the Polaris supercomputer. The output 
of these models was then post-processed 
in 147 seconds using 128 CPU nodes in the ThetaKNL 
supercomputer. 

\begin{figure*}[!htbp]
   \centering
   \includegraphics[width=\textwidth]{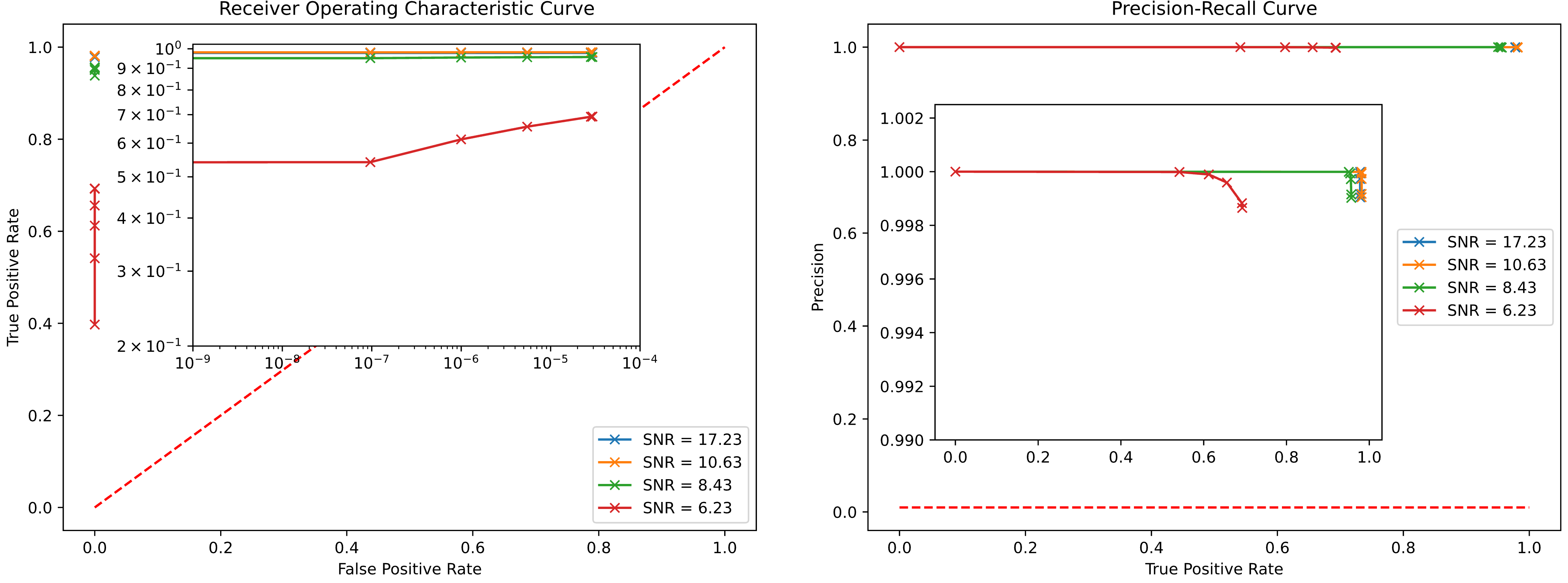}
       \caption{Receiver Operating Characteristic curve (left), and Precision Recall curve (right) for an ensemble of 3 classifiers and 2 mass predictors. Results are shown for a year's long test set in 
       which we injected 300,000 higher order mode waveforms.}
   \label{fig:ROC_PR}
\end{figure*}

\noindent \textbf{Figures of merit.} We used the post-processed data to compute the ROC AUC and the PR AUC 
using our ensemble of three AI classifiers with and 
without total mass predictors. To post-process 
noise triggers identified by our AI 
classifiers, we used the total mass predictors. 
In practice, 
a filter is applied that considers total masses greater 
than $5M_\odot$ as potential events and disregards 
those below as noise. This threshold is significantly 
beneath the range of total masses used during training. 
The final mass estimation is derived by averaging the 
outputs of our 2 total mass predictors. 
To create the ROC curve, we computed the 
true positive rate against the false positive rate as 
estimated from the output of our AI ensembles when 
they process a year-long test set 
for the HLV network. We injected 
300,000 modeled waveforms in this test set, and 
quantified how accurately our AI ensembles correctly 
identified injected 
signals, and discarded other noise anomalies. To 
produce the PR curve, we consider that $\textrm{PR}= \textrm{TP}/\left(\textrm{TP}+\textrm{FP}\right)$, 
where $\textrm{TP}$ and $\textrm{FP}$ stand for 
True Positives and False Positives, respectively. 
Results 
for the ROC AUC and the PR AUC are presented in 
Figure~\ref{fig:ROC_PR}. We present the false 
positive results with different threshold ${\cal H}$ 
in Table~\ref{tab:fpr}.

A positive signal detection is only declared when there is unanimous agreement among all AI classifiers, else it is classified as a noise anomaly or pure noise. This method of consensus aids in minimizing the FPR, which is crucial for practical applications, ensuring that the detected signals are indeed representative of true events. This 
approach also provides a rapid estimate of 
the total mass of potential gravitational wave sources.

\begin{table*}[!htbp]
\centering
  \begin{tabular}{ccccccc}
    \toprule
    \multicolumn{1}{c}{Threshold} &
    \multicolumn{2}{c}{Ensemble w/o mass estimation} &
    \multicolumn{2}{c}{Ensemble w/ mass estimation} \\
    ${\cal H}$ & {\# FP/year} & {FPR} & {\# FP/year} & {FPR} \\
    \cmidrule(lr){1-1} \cmidrule(lr){2-3}\cmidrule(lr){4-5}
    0.999999 & 3   & 9.65e-8  & 2   & 6.43e-8 \\
    0.99999  & 31  & 9.97e-7  & 19  & 6.11e-7 \\
    0.9999   & 170 & 5.47e-6  & 80  & 2.57e-6 \\
    0.99     & 876 & 2.82e-5  & 244 & 7.84e-6 \\
    0.9      & 913 & 2.94e-5  & 281 & 9.03e-6 \\ 
    \bottomrule
  \end{tabular}
\caption{False positive results for different height thresholds, ${\cal H}$, and different model ensembles.}
\label{tab:fpr}
\end{table*}

\noindent \textbf{A note on AI inference.} Though we 
have developed a framework for hyper efficient, distributed 
AI inference, our open source AI ensemble may be 
readily used to process an entire month of data within 
1 hour using a 
single NVIDIA V100 GPU.

\section{Conclusions}

We introduced a novel AI ensemble of 
classifiers and predictors that consist 
of spatiotemporal-graph models that can 
process gravitational wave data from a three 
detector network faster than real-time with a 
single GPU. 
To the best of our knowledge, these AI models 
are the first of their kind in the literature that 
have been trained to search for and find higher 
order gravitational wave modes that describe 
quasi-circular, spinning, non-precessing binary 
black hole mergers. 

We quantified the performance of these 
models by processing a year's long test set within 
5 minutes using supercomputers housed at the 
\href{https://www.alcf.anl.gov}{ALCF}, 
and found 
that they provide state-of-the-art signal detection 
accuracy, and only two false positives per year of 
searched data. The software and computational 
resources we have developed in this article will 
be used in future work to train similar 
AI models using gravitational wave data 
from the Gravitational Wave Open 
Science Center~\cite{KAGRA:2023pio}. 
We also plan on improving the scalability of 
our distributed training methods to further 
reduce time-to-solution in high performance computing 
platforms.

\begin{ack}
This work was supported by Laboratory 
Directed Research and Development (LDRD) funding 
from Argonne National Laboratory, provided by the 
Director, Office 
of Science, of the U.S. Department of Energy under 
Contract No. DE-AC02-06CH11357.
E.A.H. and M.T. gratefully acknowledge 
support from National Science Foundation awards 
OAC-1931561 and OAC-2209892. This work was supported in part by the Diaspora project funded by the U.S.\ Department of Energy under Contract DE-AC02-06CH11357.  This research used 
resources of the Argonne 
Leadership Computing Facility, which is a 
DOE Office of Science User Facility supported 
under Contract DE-AC02-06CH11357. 
This research used the Delta advanced computing 
and data resource which is 
supported by the National Science Foundation (award 
OAC 2005572) and the State of Illinois. Delta is a 
joint effort of the University of Illinois at
Urbana-Champaign and its National Center for 
Supercomputing Applications. 
\end{ack}

\bibliographystyle{ieeetr}
\bibliography{references}

\begin{thebibliography}{10}

\bibitem{DeepLearning}
Y.~LeCun, Y.~Bengio, and G.~Hinton, ``Deep learning,'' {\em Nature}, vol.~521,
  no.~7553, pp.~436--444, 2015.

\bibitem{2022NatRP761K}
M.~{Krenn}, R.~{Pollice}, S.~Y. {Guo}, M.~{Aldeghi}, A.~{Cervera-Lierta},
  P.~{Friederich}, G.~{dos Passos Gomes}, F.~{H{\"a}se}, A.~{Jinich},
  A.~{Nigam}, Z.~{Yao}, and A.~{Aspuru-Guzik}, ``{On scientific understanding
  with artificial intelligence},'' {\em Nature Reviews Physics}, vol.~4,
  pp.~761--769, Dec. 2022.

\bibitem{Davies2021AdvancingMB}
A.~Davies, P.~Veličković, L.~Buesing, S.~Blackwell, D.~Zheng, N.~Tomašev,
  R.~Tanburn, P.~Battaglia, C.~Blundell, A.~Juhász, M.~Lackenby,
  G.~Williamson, D.~Hassabis, and P.~Kohli, ``Advancing mathematics by guiding
  human intuition with ai,'' {\em Nature}, vol.~600, pp.~70--74, 12 2021.

\bibitem{Nat_Rev_2019_Huerta}
E.~A. {Huerta}, G.~{Allen}, I.~{Andreoni}, J.~M. {Antelis}, E.~{Bachelet},
  G.~B. {Berriman}, F.~B. {Bianco}, R.~{Biswas}, M.~{Carrasco Kind},
  K.~{Chard}, M.~{Cho}, P.~S. {Cowperthwaite}, Z.~B. {Etienne}, M.~{Fishbach},
  F.~{Forster}, D.~{George}, T.~{Gibbs}, M.~{Graham}, W.~{Gropp}, R.~{Gruendl},
  A.~{Gupta}, R.~{Haas}, S.~{Habib}, E.~{Jennings}, M.~W.~G. {Johnson},
  E.~{Katsavounidis}, D.~S. {Katz}, A.~{Khan}, V.~{Kindratenko}, W.~T.~C.
  {Kramer}, X.~{Liu}, A.~{Mahabal}, Z.~{Marka}, K.~{McHenry}, J.~M. {Miller},
  C.~{Moreno}, M.~S. {Neubauer}, S.~{Oberlin}, A.~R. {Olivas}, D.~{Petravick},
  A.~{Rebei}, S.~{Rosofsky}, M.~{Ruiz}, A.~{Saxton}, B.~F. {Schutz},
  A.~{Schwing}, E.~{Seidel}, S.~L. {Shapiro}, H.~{Shen}, Y.~{Shen}, L.~P.
  {Singer}, B.~M. {Sipocz}, L.~{Sun}, J.~{Towns}, A.~{Tsokaros}, W.~{Wei},
  J.~{Wells}, T.~J. {Williams}, J.~{Xiong}, and Z.~{Zhao}, ``{Enabling
  real-time multi-messenger astrophysics discoveries with deep learning},''
  {\em Nature Reviews Physics}, vol.~1, pp.~600--608, Oct. 2019.

\bibitem{cuoco_review}
E.~{Cuoco}, J.~{Powell}, M.~{Cavagli{\`a}}, K.~{Ackley}, M.~{Bejger},
  C.~{Chatterjee}, M.~{Coughlin}, S.~{Coughlin}, P.~{Easter}, R.~{Essick},
  H.~{Gabbard}, T.~{Gebhard}, S.~{Ghosh}, L.~{Haegel}, A.~{Iess}, D.~{Keitel},
  Z.~{Marka}, S.~{Marka}, F.~{Morawski}, T.~{Nguyen}, R.~{Ormiston},
  M.~{Puerrer}, M.~{Razzano}, K.~{Staats}, G.~{Vajente}, and D.~{Williams},
  ``{Enhancing Gravitational-Wave Science with Machine Learning},'' {\em Mach.
  Learn. Sci. Tech.}, vol.~2, no.~1, p.~011002, 2021.

\bibitem{huerta_book}
E.~A. {Huerta} and Z.~{Zhao}, ``{Advances in Machine and Deep Learning for
  Modeling and Real-Time Detection of Multi-messenger Sources},'' in {\em
  Handbook of Gravitational Wave Astronomy}, p.~47, Springer, Singapore, 2021.

\bibitem{huerta_nat_ast}
E.~A. {Huerta}, A.~{Khan}, X.~{Huang}, M.~{Tian}, M.~{Levental}, R.~{Chard},
  W.~{Wei}, M.~{Heflin}, D.~S. {Katz}, V.~{Kindratenko}, D.~{Mu},
  B.~{Blaiszik}, and I.~{Foster}, ``{Accelerated, scalable and reproducible
  AI-driven gravitational wave detection},'' {\em Nature Astronomy}, vol.~5,
  pp.~1062--1068, July 2021.

\bibitem{Abbott2020}
B.~P. Abbott, R.~Abbott, T.~D. Abbott, S.~Abraham, F.~Acernese, K.~Ackley,
  C.~Adams, V.~B. Adya, C.~Affeldt, M.~Agathos, K.~Agatsuma, N.~Aggarwal, O.~D.
  Aguiar, L.~Aiello, A.~Ain, P.~Ajith, T.~Akutsu, G.~Allen, A.~Allocca, M.~A.
  Aloy, P.~A. Altin, A.~Amato, A.~Ananyeva, S.~B. Anderson, W.~G. Anderson,
  M.~Ando, S.~V. Angelova, S.~Antier, S.~Appert, K.~Arai, K.~Arai, Y.~Arai,
  S.~Araki, A.~Araya, M.~C. Araya, J.~S. Areeda, M.~Ar{\`e}ne, N.~Aritomi,
  N.~Arnaud, K.~G. Arun, S.~Ascenzi, G.~Ashton, Y.~Aso, S.~M. Aston, P.~Astone,
  F.~Aubin, P.~Aufmuth, K.~AultONeal, C.~Austin, V.~Avendano, A.~Avila-Alvarez,
  S.~Babak, P.~Bacon, F.~Badaracco, M.~K.~M. Bader, S.~W. Bae, Y.~B. Bae,
  L.~Baiotti, R.~Bajpai, P.~T. Baker, F.~Baldaccini, G.~Ballardin, S.~W.
  Ballmer, S.~Banagiri, J.~C. Barayoga, S.~E. Barclay, B.~C. Barish, D.~Barker,
  K.~Barkett, S.~Barnum, F.~Barone, B.~Barr, L.~Barsotti, M.~Barsuglia,
  D.~Barta, J.~Bartlett, M.~A. Barton, I.~Bartos, R.~Bassiri, A.~Basti,
  M.~Bawaj, J.~C. Bayley, M.~Bazzan, B.~B{\'e}csy, M.~Bejger, I.~Belahcene,
  A.~S. Bell, D.~Beniwal, B.~K. Berger, G.~Bergmann, S.~Bernuzzi, J.~J. Bero,
  C.~P.~L. Berry, D.~Bersanetti, A.~Bertolini, J.~Betzwieser, R.~Bhandare,
  J.~Bidler, I.~A. Bilenko, S.~A. Bilgili, G.~Billingsley, J.~Birch, R.~Birney,
  O.~Birnholtz, S.~Biscans, S.~Biscoveanu, A.~Bisht, M.~Bitossi, M.~A.
  Bizouard, J.~K. Blackburn, C.~D. Blair, D.~G. Blair, R.~M. Blair, S.~Bloemen,
  N.~Bode, M.~Boer, Y.~Boetzel, G.~Bogaert, F.~Bondu, E.~Bonilla, R.~Bonnand,
  P.~Booker, B.~A. Boom, C.~D. Booth, R.~Bork, V.~Boschi, S.~Bose, K.~Bossie,
  V.~Bossilkov, J.~Bosveld, Y.~Bouffanais, A.~Bozzi, C.~Bradaschia, P.~R.
  Brady, A.~Bramley, M.~Branchesi, J.~E. Brau, T.~Briant, J.~H. Briggs,
  F.~Brighenti, A.~Brillet, M.~Brinkmann, V.~Brisson, P.~Brockill, A.~F.
  Brooks, D.~A. Brown, D.~D. Brown, S.~Brunett, A.~Buikema, T.~Bulik, H.~J.
  Bulten, A.~Buonanno, D.~Buskulic, C.~Buy, R.~L. Byer, M.~Cabero, L.~Cadonati,
  G.~Cagnoli, C.~Cahillane, J.~C. Bustillo, T.~A. Callister, E.~Calloni, J.~B.
  Camp, W.~A. Campbell, M.~Canepa, K.~Cannon, K.~C. Cannon, H.~Cao, J.~Cao,
  E.~Capocasa, F.~Carbognani, S.~Caride, M.~F. Carney, G.~Carullo, J.~C. Diaz,
  C.~Casentini, S.~Caudill, M.~Cavagli{\`a}, F.~Cavalier, R.~Cavalieri,
  G.~Cella, P.~Cerd{\'a}-Dur{\'a}n, G.~Cerretani, E.~Cesarini, O.~Chaibi,
  K.~Chakravarti, S.~J. Chamberlin, M.~Chan, M.~L. Chan, S.~Chao, P.~Charlton,
  E.~A. Chase, E.~Chassande-Mottin, D.~Chatterjee, M.~Chaturvedi,
  K.~Chatziioannou, B.~D. Cheeseboro, C.~S. Chen, H.~Y. Chen, K.~H. Chen,
  X.~Chen, Y.~Chen, Y.~R. Chen, H.~P. Cheng, C.~K. Cheong, H.~Y. Chia,
  A.~Chincarini, A.~Chiummo, G.~Cho, H.~S. Cho, M.~Cho, N.~Christensen, H.~Y.
  Chu, Q.~Chu, Y.~K. Chu, S.~Chua, K.~W. Chung, S.~Chung, G.~Ciani, A.~A.
  Ciobanu, R.~Ciolfi, F.~Cipriano, A.~Cirone, F.~Clara, J.~A. Clark,
  P.~Clearwater, F.~Cleva, C.~Cocchieri, E.~Coccia, P.~F. Cohadon, D.~Cohen,
  R.~Colgan, M.~Colleoni, C.~G. Collette, C.~Collins, L.~R. Cominsky,
  M.~Constancio, L.~Conti, S.~J. Cooper, P.~Corban, T.~R. Corbitt,
  I.~Cordero-Carri{\'o}n, K.~R. Corley, N.~Cornish, A.~Corsi, S.~Cortese, C.~A.
  Costa, R.~Cotesta, M.~W. Coughlin, S.~B. Coughlin, J.~P. Coulon, S.~T.
  Countryman, P.~Couvares, P.~B. Covas, E.~E. Cowan, D.~M. Coward, M.~J.
  Cowart, D.~C. Coyne, R.~Coyne, J.~D.~E. Creighton, T.~D. Creighton, J.~Cripe,
  M.~Croquette, S.~G. Crowder, T.~J. Cullen, A.~Cumming, L.~Cunningham,
  E.~Cuoco, T.~D. Canton, G.~D{\'a}lya, S.~L. Danilishin, S.~D'Antonio,
  K.~Danzmann, A.~Dasgupta, C.~F. Da~Silva~Costa, L.~E.~H. Datrier, V.~Dattilo,
  I.~Dave, M.~Davier, D.~Davis, E.~J. Daw, D.~DeBra, M.~Deenadayalan,
  J.~Degallaix, M.~De~Laurentis, S.~Del{\'e}glise, W.~D. Pozzo, L.~M. DeMarchi,
  N.~Demos, T.~Dent, R.~De~Pietri, J.~Derby, R.~De~Rosa, C.~De~Rossi,
  R.~DeSalvo, O.~de~Varona, S.~Dhurandhar, M.~C. D{\'\i}az, T.~Dietrich, and
  L.~D. Fiore, ``Prospects for observing and localizing gravitational-wave
  transients with advanced ligo, advanced virgo and kagra,'' {\em Living
  Reviews in Relativity}, vol.~23, no.~1, p.~3, 2020.

\bibitem{pdsnoise}
{LIGO Document Control Center Portal}, ``{Noise curves used for Simulations in
  the update of the Observing Scenarios Paper},'' 2022.
\newblock \url{https://dcc.ligo.org/LIGO-T2000012/public}.

\bibitem{KAGRA:2023pio}
R.~Abbott {\em et~al.}, ``{Open Data from the Third Observing Run of LIGO,
  Virgo, KAGRA, and GEO},'' {\em Astrophys. J. Suppl.}, vol.~267, no.~2, p.~29,
  2023.

\bibitem{2021PhRvD.103j4056P}
G.~{Pratten}, C.~{Garc{\'\i}a-Quir{\'o}s}, M.~{Colleoni}, A.~{Ramos-Buades},
  H.~{Estell{\'e}s}, M.~{Mateu-Lucena}, R.~{Jaume}, M.~{Haney}, D.~{Keitel},
  J.~E. {Thompson}, and S.~{Husa}, ``{Computationally efficient models for the
  dominant and subdominant harmonic modes of precessing binary black holes},''
  {\em Phys. Rev. D}, vol.~103, p.~104056, May 2021.

\bibitem{pycbc_library}
A.~H. {Nitz}, I.~W. {Harry}, D.~A. {Brown}, C.~M. {Biwer}, J.~L. {Willis},
  T.~{Dal Canton}, C.~D. {Capano}, L.~P. {Pekowsky}, T.~{Dent}, A.~R.
  {Williamson}, G.~S. {Davies}, S.~{De}, , M.~{Cabero}, B.~{Machenschalk},
  P.~{Kumar}, S.~{Reyes}, D.~{MacLeod}, D.~{Finstad}, F.~{Pannarale},
  T.~{Massinger}, S.~{Kumar}, M.~{Tapai}, L.~{Singer}, S.~{Khan},
  S.~{Fairhurst}, A.~{Nielsen}, and S.~{Singh}, ``{PyCBC. Free and open
  software to study gravitational waves},'' 2021.

\bibitem{2023arXiv230615728T}
M.~{Tian}, E.~A. {Huerta}, and H.~{Zheng}, ``{Physics-inspired
  spatiotemporal-graph AI ensemble for gravitational wave detection},'' {\em
  arXiv e-prints}, p.~arXiv:2306.15728, June 2023.

\bibitem{2016wavenet}
A.~{van den Oord}, S.~{Dieleman}, H.~{Zen}, K.~{Simonyan}, O.~{Vinyals},
  A.~{Graves}, N.~{Kalchbrenner}, A.~{Senior}, and K.~{Kavukcuoglu},
  ``{WaveNet: A Generative Model for Raw Audio},'' in {\em 9th ISCA Speech
  Synthesis Workshop}, pp.~125--125, 2016.

\bibitem{bruna2013spectral}
J.~Bruna, W.~Zaremba, A.~Szlam, and Y.~LeCun, ``Spectral networks and locally
  connected networks on graphs,'' {\em arXiv preprint arXiv:1312.6203}, 2013.

\bibitem{gilmer2017neural}
J.~Gilmer, S.~S. Schoenholz, P.~F. Riley, O.~Vinyals, and G.~E. Dahl, ``Neural
  message passing for quantum chemistry,'' in {\em International conference on
  machine learning}, pp.~1263--1272, PMLR, 2017.

\bibitem{NIPS2017_5dd9db5e}
W.~Hamilton, Z.~Ying, and J.~Leskovec, ``Inductive representation learning on
  large graphs,'' in {\em Advances in Neural Information Processing Systems}
  (I.~Guyon, U.~V. Luxburg, S.~Bengio, H.~Wallach, R.~Fergus, S.~Vishwanathan,
  and R.~Garnett, eds.), vol.~30, Curran Associates, Inc., 2017.

\bibitem{velickovic2017graph}
P.~Velickovic, G.~Cucurull, A.~Casanova, A.~Romero, P.~Lio, Y.~Bengio, {\em
  et~al.}, ``Graph attention networks,'' {\em stat}, vol.~1050, no.~20,
  pp.~10--48550, 2017.

\bibitem{xu2018powerful}
K.~Xu, W.~Hu, J.~Leskovec, and S.~Jegelka, ``How powerful are graph neural
  networks?,'' {\em arXiv preprint arXiv:1810.00826}, 2018.

\bibitem{sergeev2018horovod}
A.~Sergeev and M.~D. Balso, ``Horovod: fast and easy distributed deep learning
  in {TensorFlow},'' {\em arXiv preprint arXiv:1802.05799}, 2018.

\bibitem{you2019large}
Y.~You, J.~Li, S.~Reddi, J.~Hseu, S.~Kumar, S.~Bhojanapalli, X.~Song,
  J.~Demmel, K.~Keutzer, and C.-J. Hsieh, ``Large batch optimization for deep
  learning: Training bert in 76 minutes,'' {\em arXiv preprint
  arXiv:1904.00962}, 2019.

\bibitem{2020SciPy-NMeth}
P.~Virtanen, R.~Gommers, T.~E. Oliphant, M.~Haberland, T.~Reddy, D.~Cournapeau,
  E.~Burovski, P.~Peterson, W.~Weckesser, J.~Bright, S.~J. {van der Walt},
  M.~Brett, J.~Wilson, K.~J. Millman, N.~Mayorov, A.~R.~J. Nelson, E.~Jones,
  R.~Kern, E.~Larson, C.~J. Carey, {\.I}.~Polat, Y.~Feng, E.~W. Moore,
  J.~{VanderPlas}, D.~Laxalde, J.~Perktold, R.~Cimrman, I.~Henriksen, E.~A.
  Quintero, C.~R. Harris, A.~M. Archibald, A.~H. Ribeiro, F.~Pedregosa, P.~{van
  Mulbregt}, and {SciPy 1.0 Contributors}, ``{{SciPy} 1.0: Fundamental
  Algorithms for Scientific Computing in Python},'' {\em Nature Methods},
  vol.~17, pp.~261--272, 2020.

\end{thebibliography}

\medskip

\end{document}